# Fluctuation electromagnetic conservative–dissipative interaction and heating of two closely spaced parallel plates in relative motion. Nonrelativistic approximation.1.

G.V. Dedkov, A.A. Kyasov

*Nanoscale Physics Group, Kabardino –Balkarian State University, Nalchik, 360004,*
*Russian Federation*
*E-mail: gv_dedkov@mail.ru*

For the first time, we calculate the heating rate, attractive conservative and tangential dissipative fluctuation electromagnetic forces felt by a thick plate moving parallel to a closely spaced another plate in rest using a nonrelativistic approximation of fluctuation electrodynamics. These results can be considered as the high lights when solving general relativistic problem of the fluctuation electromagnetic interaction in configuration of two perfectly smooth parallel thick plates in relative motion.

## 1.Introduction

Vacuum attraction, friction and heat exchange of neutral nonmagnetic bodies moving with relative velocity $V$ are the well known effects of electromagnetic fluctuations. To date, however, theoretical description of many aspects of fluctuation electromagnetic interactions (FEI) has encountered with a lot of problems attracting steady growing attention (see the reviewing papers [1,2,3,4,5,6,7,8,9]). Of these one can mention the problems of dissipative (frictional) forces [2,3,6], the thermodynamics puzzles of the Lifshitz theory [1,5,6] and non –equilibrium Casimir forces [10], etc. The range of applications involving FEI is very wide and extends from atomic physics and elementary particle physics to astrophysics and cosmology. By measuring Casimir forces, for example, one can study structure of quantum vacuum and determine restrictions on the magnitude of hypothetical long-range forces that are corrections to Newtonian gravitational forces [1]. Under outer space conditions, FEI between dust particles and background electromagnetic radiation can play an important role in evolution of gas –dust clouds.

In general, FEI is associated with quantum and thermal fluctuations in the polarization and magnetization of condensed bodies. Calculating the spectrum of electromagnetic fluctuations for arbitrary geometry of interacting bodies poses severe difficulties. For this reason, exact analytical or numerical solutions of the problems related with FEI (if these solutions exist) are of fundamental physical importance. One of that kind problems has been recently solved and reviewed in detail in our papers [7,9]. It corresponds to the geometrical configuration "small spherical particle –plate", further referred to as configuration "1" (see Fig.1(a)). But historically, since pioneering works by Casimir and Lifshitz [11,12], the most widely used configuration in calculations FEI was regarded another one, corresponding to smooth featureless parallel plates divided by a vacuum gap of width $z$ (Fig.1(b)). In what follows this configuration is called "2"[1]. Note that to date, contrary to configuration 1, a strict solution of the problem FEI in configuration 2 in relativistic statement and out of thermal equilibrium is not yet obtained. Different aspects of this matter have been discussed in [2,3,6] and revealed many contradictions between the results of different authors. Moreover, even in the nonrelativistic and nonretarded case several points of this problem seem to be insufficiently clear. For configuration 1, on the other hand, despite the problem statement is much more transparent and allows to obtain an explicit solution, its fundamental significance in the theory of FEI has not been properly appreciated to date.

The aim of this paper is to work out the existing drawbacks in configuration 2 using our exact solution relevant to configuration 1. We show that, starting from this one, we are able to get in an unambiguous way a set of expressions for the conservative –dissipative fluctuation electromagnetic forces and rate of heating in a system of two closely spaced parallel plates in relative motion. We also discuss the recent results obtained in [13] and feature its inconsistency points. We argue that closed solution of general relativistic problem in configuration 2 still presents a challenge for further investigations.

## 2. Configuration small spherical particle –plate. Basic relativistic results

We start with exact relativistic expressions which we obtained in configuration small particle – plate for the conservative (dissipative) forces and rate of particle heating caused by FEI [7,9].The particle was modeled by a sphere of radius $R$, and the dipole approximation $R/z << 1$ was assumed, where $z$ is a distance between the center of the sphere and the plate. Geometry of motion of particle and the coordinate system used are shown in Fig.1(a). Eqs. (12)-(14) in Ref. [7] for the force components $F_x, F_z$ and heating rate $dQ/dt$ in the reference frame of resting plate (laboratory frame) can be simplified further by making use of an expansion of the integration domains over the wave vectors $k_x, k_y$ to full axes $(-\infty, +\infty)$, and omitting in (1) and (3) the terms related with interaction of a particle with vacuum background modes (the omitted terms are independent of $z$):

$$F_x = -\frac{\hbar\gamma}{2\pi^2}\int_0^\infty d\omega \int_{-\infty}^{+\infty} dk_x \int_{-\infty}^{+\infty} dk_y k_x \alpha''(\gamma(\omega^+)) \operatorname{Im}\left[\frac{\exp(-2q_0 z)}{q_0} R(\omega,\mathbf{k})\right]\cdot$$
$$\cdot\left[\coth\left(\frac{\hbar\omega}{2k_B T_2}\right) - \coth\left(\frac{\gamma\hbar\omega^+}{2k_B T_1}\right)\right] \tag{1}$$

$$F_z = -\frac{\hbar\gamma}{2\pi^2}\int_0^\infty d\omega \int_{-\infty}^{+\infty} dk_x \int_{-\infty}^{+\infty} dk_y \cdot$$
$$\cdot\left\{\begin{array}{l} \alpha''(\gamma(\omega^+))\operatorname{Re}\left[\exp(-2q_0 z)R(\omega,\mathbf{k})\right]\coth\left(\dfrac{\gamma\hbar\omega^+}{2k_B T_1}\right) + \\[2mm] + \alpha'(\gamma(\omega^+))\operatorname{Im}\left[\exp(-2q_0 z)R(\omega,\mathbf{k})\right]\coth\left(\dfrac{\hbar\omega}{2k_B T_2}\right) \end{array}\right\} \tag{2}$$

$$\frac{dQ}{dt} = \frac{\hbar\gamma}{2\pi^2}\int_0^\infty d\omega \int_{-\infty}^{+\infty} dk_x \int_{-\infty}^{+\infty} dk_y (\omega + k_x V)\alpha''(\gamma(\omega^+)) \operatorname{Im}\left[\frac{\exp(-2q_0 z)}{q_0} R(\omega,\mathbf{k})\right]\cdot$$
$$\cdot\left[\coth\left(\frac{\hbar\omega}{2k_B T_2}\right) - \coth\left(\frac{\gamma\hbar\omega^+}{2k_B T_1}\right)\right] \tag{3}$$

$$\Delta_e(\omega) = \frac{q_0\varepsilon(\omega) - q}{q_0\varepsilon(\omega) + q}, \quad \Delta_m(\omega) = \frac{q_0\mu(\omega) - q}{q_0\mu(\omega) + q}, \quad q = \left(k^2 - (\omega^2/c^2)\varepsilon(\omega)\mu(\omega)\right)^{1/2}$$
$$q_0 = (k^2 - \omega^2/c^2)^{1/2}, \ k^2 = |\mathbf{k}|^2 = k_x^2 + k_y^2, \beta = V/c, \gamma = (1-\beta^2)^{-1/2}, \omega^+ = \omega + k_x V \tag{4}$$

$$R(\omega,\mathbf{k}) = \Delta_e(\omega)\left[2(k^2 - k_x^2\beta^2)(1 - \omega^2/k^2 c^2) + (\omega + k_x V)^2/c^2\right] +$$
$$+ \Delta_m(\omega)\left[2k_y^2\beta^2(1 - \omega^2/k^2 c^2) + (\omega + k_x V)^2/c^2\right] \tag{5}$$

---

[1] Note that in Ref.[9] we used an opposite way of numeration : "1" denoted configuration "2" in this work and vice versa.

Moreover, $\varepsilon(\omega)$ and $\mu(\omega)$ are the frequency –dependent dielectric and magnetic permittivities of the plate material, $\alpha(\omega)$ is the frequency –dependent dipole electric polarizability of the particle. The contributions related with dipole magnetic polarizability of the particle have to be expressed by exactly the same equations (1)-(4) with a simple replacement $e \leftrightarrow m$ in (5) and assuming that $\alpha(\omega)$ denotes magnetic polarizability. One primed and double primed quantities in (1)-(3) represent the corresponding real and imagine parts. Also, it is assumed that the plate and surrounding vacuum background are in the state of thermal equilibrium at temperature $T_2$ (see Fig.1(a)).

Comparing Eqs.(1)-(3) with Eqs.(12)-(14) in Ref. [7], one should take into account the necessary relations

$$\int\limits_{-\infty}^{+\infty}\int\limits_{-\infty}^{+\infty} d^2k = \int\limits_{k>\omega/c} d^2k + \int\limits_{k<\omega/c} d^2k \ , \ q_0 = (k^2 - \omega^2/c^2)^{1/2}, \ \tilde{q}_0 = (\omega^2/c^2 - k^2)^{1/2}, \ q_0 \to -i\tilde{q}_0 \qquad (6)$$

An important advantage of the formulae (1)-(3) in comparison with (12)-(14) in Ref.[7] is that the contributions from evanescent modes ($k > \omega/c$) and from wave–modes ($k < \omega/c$) are combined into a single integral term, since the electromagnetic modes of both types come into all resulting formulae in a similar way, being related by analytic transformation. This fact proves to be of principal value.

In what follows we are concerned with the nonrelativistic ($\beta = V/c \to 0$) and nonretarded ($\omega z/c \to 0$) approximation, while the sample plate is assumed to be a nonmagnetic substance. Then, making use of the above approximations in (1)-(5) yields

$$F_x = -\frac{\hbar}{\pi^2}\int\limits_0^\infty d\omega \int\limits_{-\infty}^{+\infty} dk_x \int\limits_{-\infty}^{+\infty} dk_y k k_x \exp(-2kz)\Delta''(\omega)\alpha''(\omega^+)\left[\coth\left(\frac{\hbar\omega}{2k_B T_2}\right) - \coth\left(\frac{\hbar\omega^+}{2k_B T_1}\right)\right] \qquad (7)$$

$$F_z = -\frac{\hbar}{\pi^2}\int\limits_0^\infty d\omega \int\limits_{-\infty}^{+\infty} dk_x \int\limits_{-\infty}^{+\infty} dk_y k^2 \exp(-2kz)\left[\begin{array}{l}\Delta''(\omega)\alpha'(\omega^+)\coth\left(\dfrac{\hbar\omega}{2k_B T_2}\right)+ \\[2mm] +\Delta'(\omega)\alpha''(\omega^+)\coth\left(\dfrac{\hbar\omega^+}{2k_B T_1}\right)\end{array}\right] \qquad (8)$$

$$\frac{dQ}{dt} = \frac{\hbar}{\pi^2}\int\limits_0^\infty d\omega \int\limits_{-\infty}^{+\infty} dk_x \int\limits_{-\infty}^{+\infty} dk_y k \exp(-2kz)\Delta''(\omega)\alpha''(\omega^+)\omega^+\left[\coth\left(\frac{\hbar\omega}{2k_B T_2}\right) - \coth\left(\frac{\hbar\omega^+}{2k_B T_1}\right)\right] \qquad (9)$$

$$\Delta_e(\omega) = \Delta(\omega) = \frac{\varepsilon(\omega)-1}{\varepsilon(\omega)+1} \ , \ \Delta_m(\omega) = 0, \ k = (k_x^2 + k_y^2)^{1/2} \qquad (10)$$

Formulae (7)-(9) have been firstly derived in our papers [2,3,14] when solving the same nonrelativistic problem, and later in [15,16] using a relativistic statement and the limit $c \to \infty$. In total, we see that formulae (7)-(9) describe the contribution of evanescent surface modes ($k > \omega/c$) in the fluctuation electromagnetic force and rate of heating (cooling) of a moving nonrelativistic particle at different temperature of the particle ($T_1$) and the sample surface ($T_2$). A contribution from surface wave modes ($k < \omega/c$) in the limit $c \to \infty$ goes to zero.

It is worth noticing that expressions (7)-(9) are valid irrespectively of the state of thermal equilibrium in a system "vacuum background –plate" : $T_2 = T_3$ or $T_2 \neq T_3$ , where $T_3$ is the background temperature or the temperature of distant environment bodies. The same statement holds also for evanescent wave contributions in the relativistic formulae (1)-(3) [7,9]. But, contrary to that, the terms related with surface wave modes in (1) –(3) turn out to be essentially dependent on whether the condition of local thermal equilibrium $T_2 = T_3$ is fulfilled, or not [17].

The above expressions (1)-(3) and (7)-(9) can be considered as the referring basic high lights characterizing the involved geometrical configuration 1. In the next section we show that, starting from Eqs. (7)-(9), it is possible to obtain in an unambiguous way the expressions for the conservative –dissipative fluctuation electromagnetic forces and rates of heating in configuration 2 (Fig.1(b)).

### 3. A system of two parallel plates in relative motion

Configuration of two perfectly plane infinite parallel plates in rest separated by a vacuum gap of width $z$ is the standard Casimir configuration which is used in calculations of conservative fluctuation electromagnetic forces [1,5,10,11,12]. Configuration 1, to date, has been considered by many authors as less important, secondary one, because the Casimir–Polder force between a small particle (an atom) and a plate can be calculated in the limit of rarified material for one of the interacting plates via the Lifshitz formula for the Casimir force between two parallel plates [18]. In this case one must employ the relation $\varepsilon_1(\omega) - 1 = 4\pi n_1 \alpha_1(\omega) \to 0$ ( $n_1$ is the atomic density of an upper plate) while the correspondence rule reads [10]

$$F_z^{(1)} = -\frac{1}{n_1 S} \frac{dF_z^{(2)}(l)}{dl}\Big|l = z \tag{11}$$

with $F_z^{(2)}(l)/S$ being the Casimir –Lifshitz attraction force per unit area of two parallel plates divided by a gap of width $l$ . The force in the left hand side of (11) denotes the Casimir–Polder force applied to a small particle (an atom), which is located a distance $z$ apart from the plate. Quite recently, configuration (2) and prescription (11) have been used both in calculations of normal and lateral forces applied to a moving small particle [13]. The necessary force projections $F^{(2)}{}_{x,z}$ were calculated using a relativistic modification of the Lifshitz theory provided the needed components of the Maxwell stress tensor are known.

However, a clear correspondence between the theories in configurations 1 and 2 (Fig.1(a,b)) is not so trivial, because the original Lifshitz theory was developed under condition of total thermodynamic equilibrium [12]. On the contrary, the problem statements shown in Fig.1 are quite different from that one: both systems under study are out of thermal and mechanical equilibrium simultaneously.

In this relation, two essential points are worthy of attention: i) when and under what conditions the prescriptions like (11) are valid ? ii) how should we use them to obtain correct results in configuration 2, if we start from Eqs. (1)-(3) or from (8)-(9) ? It seems to be quite natural to believe that results obtained when coming from configuration 1 to configuration 2 and vice versa should be interlinked. Below we aim to demonstrate that such unambiguous interconnection exists in the nonrelativistic case.

First, using Eq.(9) let us write down the expression for the attraction force in configuration 1 at $V = 0, T_1 = T_2 = T$ :

$$F_z = -\frac{\hbar}{\pi^2} \int\limits_0^\infty d\omega \int\limits_{-\infty}^{+\infty} dk_x \int\limits_{-\infty}^{+\infty} dk_y k^2 \exp(-2kz) \left[ \begin{array}{l} \Delta''(\omega)\alpha'(\omega)\coth\left(\dfrac{\hbar\omega}{2k_BT}\right) + \\ + \Delta'(\omega)\alpha''(\omega)\coth\left(\dfrac{\hbar\omega}{2k_BT}\right) \end{array} \right] \qquad (12)$$

Comparing (12) with (8) one sees that transition to dynamic and thermal situation out of equilibrium is performed via the transformations ( $\omega^+ = \omega + k_x V$ )

$$\Delta''(\omega)\coth\left(\frac{\hbar\omega}{2k_BT}\right) \rightarrow \Delta''(\omega)\coth\left(\frac{\hbar\omega}{2k_BT_2}\right),$$

$$\alpha''(\omega)\coth\left(\frac{\hbar\omega}{2k_BT}\right) \rightarrow \alpha''(\omega^+)\coth\left(\frac{\hbar\omega^+}{2k_BT_1}\right), \qquad (13)$$

$$\alpha'(\omega),\alpha''(\omega) \rightarrow \alpha'(\omega^+),\alpha''(\omega^+)$$

On the other hand, comparing (7) and (8) shows that the tangential force $F_x$ is obtained from the normal force $F_z$ with the help of the transformations

$$d^2kk \rightarrow d^2kk_x, \Delta''(\omega) \rightarrow \Delta''(\omega), \Delta'(\omega) \rightarrow \Delta''(\omega), \alpha'(\omega^+) \rightarrow \alpha''(\omega^+), \alpha'(\omega^+) \rightarrow -\alpha''(\omega^+) \qquad (14)$$

Furthermore, from (7) and (9) it follows that $dQ/dt$ is to be obtained from $F_x$ on using

$$d^2kk_x \rightarrow -d^2k\omega^+ \qquad (15)$$

Analogously to (11), the interrelations between the lateral forces $F_x^{(1,2)}$ and heating rates $dQ^{(1,2)}/dt$ in configurations 1 and 2 are given by

$$F_x^{(1)} = -\frac{1}{n_1 S}\frac{dF_x^{(2)}(l)}{dl}\Big|l = z \, , \, dQ^{(1)}/dt = -\frac{1}{n_1 S}\frac{d\dot{Q}^{(2)}(l)}{dl}\Big|l = z \qquad (16)$$

where $F_x^{(2)}(l)$ and $\dot{Q}^{(2)}(l) = dQ^{(2)}/dt$ are the applied lateral force and heating rate of a moving plate in configuration 2 in the limit of rarified medium. As the formulae (7)-(9) must follow from the corresponding ones in configuration 2 with the help of linear transformation $\varepsilon_1(\omega) - 1 = 4\pi n_1 \alpha_1(\omega) \to 0$, then the quantities $F_x^{(2)}(l)$, $F_z^{(2)}(l)$, $\dot{Q}^{(2)}(l)$ should be related by the relations similar to (13), (14), (15) with the replacement $\alpha(\omega) \to \Delta_1(\omega)$.

As we have done in the case of configuration 1, first let us consider an expression for the force of the nonretarded attraction (Van–der–Waals force) of two parallel plates at $V = 0, T_1 = T_2 = T$, which we rewrite in a more convenient form

$$F_z^{(2)} = -\frac{\hbar S}{4\pi^3} \int_0^{+\infty} d\omega \int_{-\infty}^{+\infty} dk_x \int_{-\infty}^{+\infty} dk_y k \frac{\exp(-2kl)}{\left|1 - \exp(-2kl)\Delta_1(\omega)\Delta_2(\omega)\right|^2} \cdot$$
$$\cdot \left[\Delta_1''(\omega)\Delta_2''(\omega)\coth\left(\hbar\omega/2k_BT\right) + \Delta_1'(\omega)\Delta_2''(\omega)\coth\left(\hbar\omega/2k_BT\right)\right] \qquad (17)$$

where $\Delta_1(\omega) = \dfrac{\varepsilon_1(\omega) - 1}{\varepsilon_1(\omega) + 1}$ and $\Delta_2(\omega) = \dfrac{\varepsilon_2(\omega) - 1}{\varepsilon_2(\omega) + 1}$, $\varepsilon_{1,2}(\omega)$ are the dielectric permittivities of the plates 1,2 respectively. Making use in (17) the transformations

$$\Delta_2''(\omega)\coth\left(\frac{\hbar\omega}{2k_BT}\right) \to \Delta_2''(\omega)\coth\left(\frac{\hbar\omega}{2k_BT_2}\right),$$
$$\Delta_1''(\omega)\coth\left(\frac{\hbar\omega}{2k_BT}\right) \to \Delta_1''(\omega^+)\coth\left(\frac{\hbar\omega^+}{2k_BT_1}\right), \qquad (18)$$
$$\Delta_1'(\omega), \Delta_1''(\omega) \to \Delta_1'(\omega^+), \Delta_1''(\omega^+),$$

we immediately obtain the expression for the attraction force between two moving parallel plates out of equilibrium in configuration 2 (Fig.1(b)):

$$F_z^{(2)} = -\frac{\hbar S}{4\pi^3} \int_0^{\infty} d\omega \int_{-\infty}^{+\infty} dk_x \int_{-\infty}^{+\infty} dk_y k \frac{\exp(-2kl)}{\left|1 - \exp(-2kl)\Delta_1(\omega^+)\Delta_2(\omega)\right|^2} \cdot$$
$$\cdot \left[\Delta_1''(\omega^+)\Delta_2'(\omega)\coth\left(\hbar\omega^+/2k_BT_1\right) + \Delta_1'(\omega^+)\Delta_2''(\omega)\coth\left(\hbar\omega/2k_BT_2\right)\right] \qquad (19)$$

Similar to (7),(8) we must perform in (19) the following transformations in order to get the involved expression for $F_x^{(2)}$ :

$$d^2kk \to d^2kk_x, \Delta_2''(\omega) \to \Delta_2''(\omega), \Delta_2'(\omega) \to \Delta_2''(\omega), \Delta_1'(\omega^+) \to \Delta_1''(\omega^+), \Delta_1''(\omega^+) \to -\Delta_1''(\omega^+) \quad (20)$$

After doing that we get

$$F_x^{(2)} = -\frac{\hbar S}{4\pi^3} \int_0^{\infty} d\omega \int_{-\infty}^{+\infty} dk_x \int_{-\infty}^{+\infty} dk_y k_x \frac{\exp(-2kl)}{\left|1 - \exp(-2kl)\Delta_1(\omega^+)\Delta_2(\omega)\right|^2} \Delta_1''(\omega^+)\Delta_2''(\omega) \cdot$$
$$\cdot \left[\coth\left(\hbar\omega/2k_BT_2\right) - \coth\left(\hbar\omega^+/2k_BT_1\right)\right] \qquad (21)$$

At last, making use the transformation $d^2kk_x \to -d^2k\omega^+$ in (21), we get the heating rate $\dot{Q}^{(2)}$:

$$\dot{Q}^{(2)} = \frac{\hbar S}{4\pi^3}\int_0^\infty d\omega \int_{-\infty}^{+\infty} dk_x \int_{-\infty}^{+\infty} dk_y \,\omega^+ \frac{\exp(-2kl)}{\left|1 - \exp(-2kl)\Delta_1(\omega^+)\Delta_2(\omega)\right|^2} \Delta_1''(\omega^+)\Delta_2''(\omega) \cdot$$
$$\cdot \left[\coth\left(\hbar\omega/2k_BT_2\right) - \coth\left(\hbar\omega^+/2k_BT_1\right)\right] \tag{22}$$

An important point is that the heat fluxes $\dot{Q}^{(1,2)}$ are produced not only due to the temperature difference between the contacting bodies, but also caused by transformation of work of lateral forces $F_x^{(1,2)}$ into heat.

## 4. A comparison with existing results by different authors

Now it is interesting to compare formulae (19), (21), (22) with the results of other authors. One of the first successive attempts to calculate the dissipative force $F_x^{(2)}$ between the perfect smooth plates has been done by Pendry [19]. However, he has considered only a simple case $T_1 = T_2 = 0$. Later, in [20] Pendry obtained the heating rate $\dot{Q}^{(2)}$ at $V = 0$ which expression has proved to be in accordance with the more general result [21] in configuration 2. Formulae (21), (22) also agree with these limiting cases. For a review of more recent calculations see [2,3]. None of these theories presented the quantities $F_x^{(2)}(l), F_z^{(2)}(l), \dot{Q}^{(2)}(l)$ in a closed set of equations similar to Eqs.(19), (21), (22).

The static $(V = 0)$ retarded Casimir–Polder and Casimir–Lifshitz forces in configurations 1,2 out of equilibrium have been calculated in [22,10]. Formulae (8) and (19) also agree with the proper results in the nonretarded limit.

Quite recently, the results for all of the quantities which we are interested in this paper have been presented by Volokitin and Persson in [13]. The authors claim that they have developed a relativistic out of thermal equilibrium modification of the Lifshitz theory in configuration 2 and, using that and the limit of one rarified body, they derived the involved formulae relevant to configuration 1. Now let us discuss the results [13] corresponding to the nonrelativistic and nonretarded case.

i)    We may assert that only Eqs. (22), (24) in Ref. [13] for the lateral forces $F_x^{(2)}$, $F_x^{(1)}$ agree with their counterparts, Eq.(21) and Eq.(7) in the present work. Eqs. (30), (31) Ref. [13] for the normal forces $F_z^{(2)}$ and $F_z^{(1)}$ are in error, because the temperature factors of different bodies must come in combination with the proper material factors, that is not the case in [13]. Thus, the results [13] are in contradiction with ours and with Ref. [10].

ii)  The rates of heating, Eqs.(34), (36) in Ref.[13] are in error, too, because their integrands contain an incorrect frequency factor $\omega$ instead of the Doppler –shifted one, $\omega^+ = \omega + k_x V$. As we shall see below, this error results in a serious unphysical consequence.

Let us rewrite Eq.(22) in a more symmetric form

$$\dot{Q}_1^{(2)} = \frac{\hbar S}{8\pi^3} \int_{-\infty}^{+\infty} dk_x \int_{-\infty}^{+\infty} dk_y \int_{-\infty}^{\infty} d\omega \frac{\omega^+ \cdot \exp(-2kl)}{\left|1 - \exp(-2kl)\Delta_1(\omega^+)\Delta_2(\omega)\right|^2} \cdot \Delta_1''(\omega^+)\Delta_2''(\omega) \cdot$$
$$\cdot \left[\coth\left(\hbar\omega / 2k_B T_2\right) - \coth\left(\hbar\omega^+ / 2k_B T_1\right)\right] \tag{23}$$

where the subscript "1" denotes the heating rate of the (moving) plate in the reference system $K$. Evidently, the heating rate $\dot{Q}_2^{(2)}$ of the second plate in its own reference system $K'$ (see Fig.1(b)) is obtained from (23) when replacing $V \to -V$ and $1 \leftrightarrow 2$. Then, by changing the frequency variable $\omega - k_x V \to \omega$, we get

$$\dot{Q}_2^{(2)} = -\frac{\hbar S}{8\pi^3} \int_{-\infty}^{+\infty} dk_x \int_{-\infty}^{+\infty} dk_y \int_{-\infty}^{\infty} d\omega \frac{\omega \cdot \exp(-2kl)}{\left|1 - \exp(-2kl)\Delta_1(\omega^+)\Delta_2(\omega)\right|^2} \cdot \Delta_1''(\omega^+)\Delta_2''(\omega) \cdot$$
$$\cdot \left[\coth\left(\hbar\omega / 2k_B T_2\right) - \coth\left(\hbar\omega^+ / 2k_B T_1\right)\right] \tag{24}$$

Furthermore, adding (23) and (24) and taking into account (21), yields

$$\dot{Q}_1^{(2)} + \dot{Q}_2^{(2)} = \frac{\hbar S}{8\pi^3} V \int_{-\infty}^{+\infty} dk_x k_x \int_{-\infty}^{+\infty} dk_y \int_{-\infty}^{\infty} d\omega \frac{\exp(-2kl)}{\left|1 - \exp(-2kl)\Delta_1(\omega^+)\Delta_2(\omega)\right|^2} \cdot \Delta_1''(\omega^+)\Delta_2''(\omega) \cdot$$
$$\cdot \left[\coth\left(\hbar\omega / 2k_B T_2\right) - \coth\left(\hbar\omega^+ / 2k_B T_1\right)\right] = -F_x^{(2)} \cdot V \tag{25}$$

where $F_x^{(2)}$ is the friction force applied to the moving plate from the resting one in the system $K$. As $F_x^{(2)} < 0$, we may write

$$\dot{Q}_1^{(2)} + \dot{Q}_2^{(2)} = |F_x| \cdot V \tag{26}$$

Eq.(26) agrees with the general relativistic expression $\dot{Q}_1^{(2)} + \dot{Q}_2^{(2)} / \gamma = |F_x| \cdot V$ [23]. Particularly, at $T_1 = T_2 = T$ in the lowest order velocity expansion, Eq.(25) reduces to

$$\dot{Q}_1^{(2)} + \dot{Q}_2^{(2)} = \frac{\hbar S}{4\pi^2} V^2 \int_0^\infty d\omega \int_0^\infty dk k^3 \frac{\exp(-2kl)\Delta_1''(\omega)\Delta_2''(\omega)}{\left|1-\exp(-2kl)\Delta_1(\omega)\Delta_2(\omega)\right|^2} \left(-\frac{\partial}{\partial\omega}\right)\coth\left(\frac{\hbar\omega}{2k_B T}\right) > 0, \qquad (27)$$

that is in accordance with the second law of thermodynamics.

Contrary to this, an equivalent of Eq.(23) in Ref. [13] in our notation takes the form

$$\dot{Q}_1^{(2)} = \frac{\hbar S}{8\pi^3} \int_{-\infty}^{+\infty} dk_x \int_{-\infty}^{+\infty} dk_y \int_{-\infty}^{\infty} d\omega \frac{\omega \cdot \exp(-2kl)}{\left|1-\exp(-2kl)\Delta_1(\omega^+)\Delta_2(\omega)\right|^2}\Delta_1''(\omega^+)\Delta_2''(\omega) \cdot$$
$$\cdot \left[\coth(\hbar\omega/2k_B T_2) - \coth(\hbar\omega^+/2k_B T_1)\right] , \ \omega^+ = \omega + k_x V \qquad (28)$$

Making use the transformations $V \to -V$ and $1 \leftrightarrow 2$ in (27) we obtain $\dot{Q}_2^{(2)}$ similar to (24). Then, summing the calculated heating rates yields

$$\dot{Q}_1^{(2)} + \dot{Q}_2^{(2)} = F_x \cdot V , \qquad (29)$$

where again $F_x < 0$, corresponding to the frictional force. Confusion of this result becomes clear in the case $T_1 = T_2 = T$ , when, in the lowest order velocity approximation, from (27) ,(28) we get:

$$\dot{Q}_1^{(2)} + \dot{Q}_2^{(2)} = -\frac{\hbar S}{4\pi^2} V^2 \int_0^\infty d\omega \int_0^\infty dk k^3 \frac{\exp(-2kl)\Delta_1''(\omega)\Delta_2''(\omega)}{\left|1-\exp(-2kl)\Delta_1(\omega)\Delta_2(\omega)\right|^2} \left(-\frac{\partial}{\partial\omega}\right)\coth\left(\frac{\hbar\omega}{2k_B T}\right) \qquad (30)$$

This implies $\dot{Q}_1^{(2)} + \dot{Q}_2^{(2)} < 0$ at $T_1 = T_2 = T$ , that proves to be in conflict with the second law of thermodynamics.

Incorrectness of the theory [13] becomes more obvious in the relativistic case. For instance, at $V = 0$ Eq. (28) in Ref. [13] strongly disagrees with the well recognized expressions for the Casimir force in configuration 2 both under and out of thermal equilibrium [12,18], while Eq. (36) in [13] (cf. also with Eq.(75) in [6]) for the retarded heating rate $\dot{Q}^{(1)}$ disagrees with [24,25]. Particularly, the factor $R(\omega, \mathbf{k})$ in (5) at $V = 0$ turns out to come into both the integrand expression for $\dot{Q}^{(1)}$ [13] (see also [6] and references therein) and in the expression for the spectral density of fluctuating electromagnetic field near a plane surface [24]. In our notations at $V \to 0$ this factor is given by

$$R(\omega, k) = (2k^2 - \omega^2/c^2)\Delta_e(\omega) + (\omega^2/c^2)\Delta_m(\omega) \qquad (31)$$

whereas in [13] the involved expression reads $2k^2\Delta_e(\omega) + (\omega^2/c^2)\Delta_m(\omega)$. This error has been reproduced in numerous works of Volokitin et. al. since their paper [26].

A crucial physical difference between the problem statements in configurations 1, 2 is that in the first one the presence of vacuum background is an important basic standpoint and, correspondingly, we have only one large body which can be in rest with respect to the background. A small particle, moving near the surface of this body (thick plate), moves simultaneously with respect to the background. In this case, the resting plate may be or may not to be in thermal equilibrium with the background radiation. This condition directly determines the structure of fluctuating electromagnetic field near the plate. For configuration 2, in contrast, the problem statement in the dynamic situation even at $T_1 = T_2$ needs to be more elaborate, because only one plate turns out to be in rest respectively to the background, whereas another plate will be braking due to the interaction with the background. In the theory of Volokitin et. al., a relation between the temperatures of the plates and that of vacuum background was either not discussed [13], or the state of thermal equilibrium is assumed [6,26]. A more detailed analysis of the relativistic situation will be given in our next paper.

## 5.Conclusion

We have obtained closed set of expressions for the conservative –dissipative forces and heating rates in a system of two parallel thick plates in relative motion in the framework of nonrelativistic approximation of fluctuation electrodynamics. The obtained formulae strictly satisfy a "correspondence principle" between the results relevant to configuration 1 (a small spherical particle above a thick plate) and configuration 2 (two infinite parallel thick plates). The results in configuration 1 are based on an exact solution of the relativistic electrodynamic problem with account of spontaneous and induced sources of fluctuations and solving the Maxwell equations subjected to the boundary conditions for the given geometry. It is shown that the derived nonrelativistic formulae for the fluctuation forces and heating rates in configurations 1,2 can be strictly obtained from one another in the limit of rarified medium for one of the plates. These results may be regarded as important high lights when solving general relativistic problem in configuration 2. Also, we have demonstrated that recently developed dynamic relativistic out of thermal equilibrium modification of the Lifshitz theory of FEI in configuration 2 [13] is erroneous as having several principal points of inconsistency. Therefore, a development of relativistic theory in configuration 2 still remains an unresolved problem.

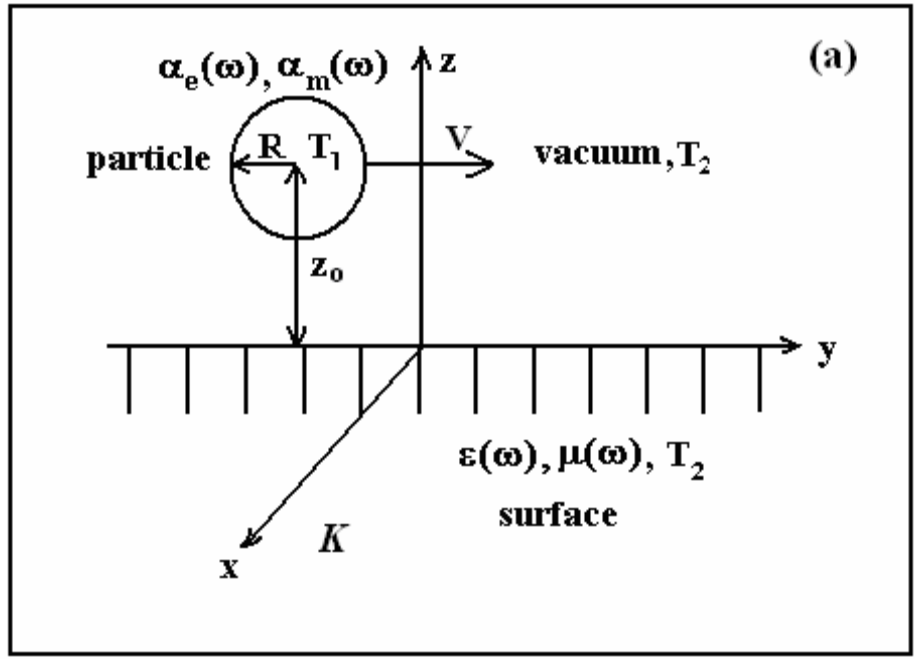

Fig.1(a)  Configuration 1. Geometry of motion of a particle and a Cartesian reference frame associated with the surface of the medium (system $K$). The Cartesian axes ($x', y', z'$) of the particle rest frame $K'$ are not shown. It is assumed that the vacuum background and surface are in thermal equilibrium

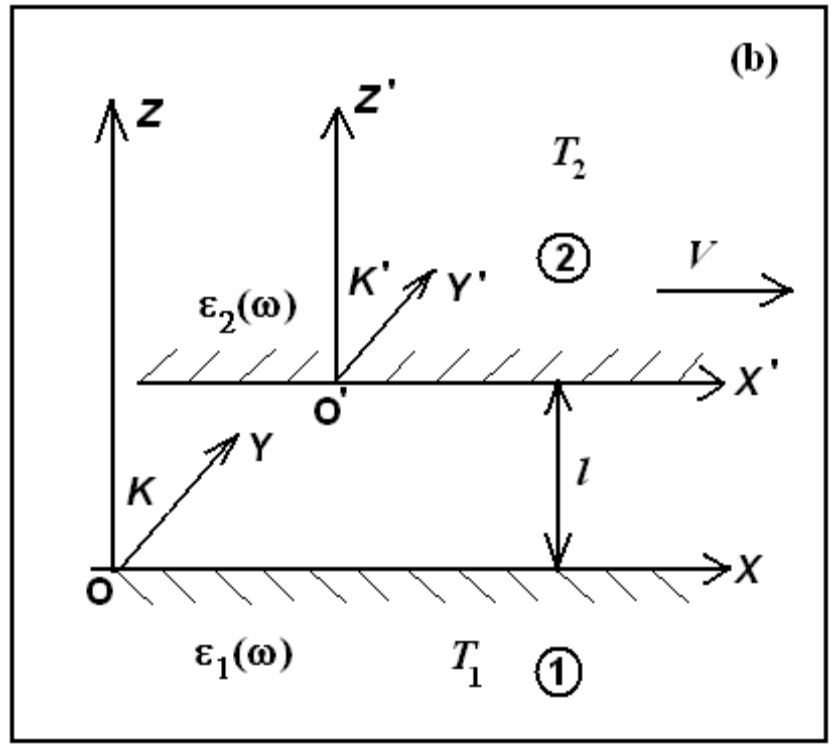

Fig.1(b) Configuration 2, corresponding to large thick plates (semi –spaces) 1 and 2 with temperatures $T_1$ and $T_2$ in the rest frame of each one, respectively. $K$ and $K'$ are the corresponding Cartesian reference frames. Surrounding vacuum background (not shown) may have, in general, the temperature $T_3$.